\documentclass[english,aps,prb,twocolumn]{revtex4}
\usepackage{amsmath}
\usepackage{amssymb}

\makeatletter

\providecommand{\tabularnewline}{\\}

\DeclareMathOperator{\arctanh}{arctanh}

\usepackage{babel}
\makeatother
\begin{document}
\begin{abstract}
Relativistic addition of velocities in one dimension, though a mainstay
of introductory physics, contributes much less physical insight than
it could. For such calculations, we propose the use of velocity factors
(two-way doppler factors). Velocities can easily, often by inspection,
be turned into velocity factors, and vice versa. Velocity factors
compose by ordinary multiplication. This simple device considerably
extends the kinds of questions that can be asked and answered in an
introductory course.
\end{abstract}

\title{Using ordinary multiplication to do relativistic velocity addition}

\author{Alma Teao Wilson}

\maketitle
Department of Mathematics\\
Brigham Young University\\
Provo, Utah, USA

\date{16 November 2006}

\section{Introduction}

Relativistic velocity addition in one dimension is a fixture of introductory
relativity. It is usually treated as a pedagogical cul-de-sac: presented,
examined, forgotten.

By a very simple transformation, often by inspection, velocities can
be converted into \emph{velocity factors,} to be defined below\emph{.}
Relativistic addition of velocities corresponds to ordinary multiplication
of velocity factors. Using the standard formula, addition of more
than two relativistic velocities quickly becomes unwieldy. Using velocity
factors, such problems can often be solved by inspection. Moreover
many problems beyond the practical reach of the usual formula are
so easily formulated using velocity factors that they, too, can be
solved by inspection.

Physically, the velocity factor is just a two-way doppler factor.
They are therefore closely related to the $k$ (a one-way doppler
factor) in Bondi's $k$-calculus \cite{Boh64,Boh64a} and also to
rapidity (the inverse hyperbolic tangent of the velocity, or logarithm
of $k$), but they are superior to either approach for deriving closed
form answers by inspection. Moreover, one can prove their properties
from the usual velocity addition formula using only elementary algebra.

\section{Relativistic velocity addition}

The one-dimensional relativistic velocity addition formula is\begin{equation}
V_{\textrm{ac}}=V_{\textrm{ab}}\boxplus V_{\textrm{bc}}=\frac{V_{\textrm{ab}}+V_{\textrm{bc}}}{1+V_{\textrm{ab}}\cdot V_{\textrm{bc}}},\label{eq:velocity-addition}\end{equation}
 where we have taken $c=1$ and where we use {}``$\boxplus$'' here
and subsequently to denote relativistic velocity addition.

A considerable advantage of this representation of the composition
of velocities is that it is clear how, for velocities small relative
to $c,$ relativistic velocity addition reduces to simple addition.
We see at once that velocity addition is commutative. Its associativity,
however, is not obvious in this representation.

It is also a pedagogical advantage, as we shall see, that when $V_{\textrm{ab}}$
and $V_{\textrm{bc}}$ are rational numbers, so is their relativitic
sum $V_{\textrm{ac}}.$

\section{A truly additive representation : rapidities}

Another representation is that in terms of \emph{rapidities} or \emph{velocity
parameters,} given by\cite{MiThWh73}\begin{equation}
\alpha=b\arctanh(\frac{V}{c}),\label{eq:rapidity-definition}\end{equation}
where $b$ is a constant. Here, it is convenient to take $b$ as unity,
but other choices can also be useful\cite{Lej80}\cite{Wia07musrel}.
We also continue to take $c=1.$

A velocity between $-1$ and $1$ becomes a rapidity between $-\infty$
and $\infty.$ This representation is monotonic increasing and invertible.
The rapidity is zero when the velocity is.

Since\begin{equation}
\tanh(\alpha_{\textrm{ab}}+\alpha_{\textrm{bc}})=\frac{\tanh(\alpha_{\textrm{ab}})+\tanh(\alpha_{\textrm{bc}})}{1+\tanh(\alpha_{\textrm{ab}})\cdot\tanh(\alpha_{\textrm{bc}})},\label{eq:tanh-of-sum}\end{equation}

the relativistic sum $V_{\textrm{ac}}=V_{\textrm{ab}}\boxplus V_{\textrm{bc}}$
yields the ordinary sum $\alpha_{\textrm{ac}}=\alpha_{\textrm{ab}}+\alpha_{\textrm{bc}}.$

The rapidity representation is manifestly commutative and manifestly
associative. Again for small velocities, the rapidity reduces to the
velocity giving correspondence with Galilean velocity addition.

Rapidity is particularly useful for integrating proper acceleration\cite{MiThWh73}.
Indeed it can be interpreted \emph{as} the integral of the proper
acceleration\cite{Lej80}: in a relativistic rocket, it is the velocity
one would calculate by multiplying the rocket's average accelerometer
reading by the elapsed time on the rocket's clock. This is the velocity
that would be imputed by an ideal Newtonian inertial guidance system.

In a companion paper, we show that rapidity can be interpreted as
the change in pitch of radiation fore and aft of the direction of
motion\cite{Wia07musrel}.

The omission of rapidities from introductory treatments of relativistic
velocity addition is puzzling. Hyperbolic tangents and their inverses
have long been available on even modest scientific calculators, so
that the result \[
V_{1}\boxplus V_{2}=\tanh\left(\arctanh V_{1}+\arctanh V_{2}\right)\]
is easy to remember and quick to compute. Nor is it a serious objection
that the calculator gives only an inexact numerical result, because
in practical situations the inexactness of computation will be dwarfed
by the inexactness of the measured values.

Nor can the reason for the omission of rapidities lie in the underlying
theory. Using velocities $V=\tanh\alpha$ but not rapidities $\alpha$
in the analytic geometry of the $x$-$t$ plane is strongly analogous
to using slope $s=\tan\theta$ but not angle $\theta$ in the analytic
geometry of the $x$-$y$ plane. One can treat the usual addition
formula for tangents as a {}``slope addition formula''\[
s_{1}\oplus s_{2}=\frac{s_{1}+s_{2}}{1-s_{1}s_{2}}.\]
 But while one can indeed formulate the analytic geometry of the Euclidean
plane using slopes and never angles, it is artificial to do so. It
is similarly artificial in relativity to use velocities and never
rapidities.

If neither theory nor practice account for this omission, perhaps
a particular kind of pedagogical convenience does. In teaching velocity
addition, it is customary to use examples and problems in which each
of the velocities to be added is a simple fraction of $c;$ their
relativistic sum is then also a fraction of $c.$ In this case, computation
using the usual velocity addition formula uses exact rational arithmetic,
which makes the examples easier to follow and the problems easier
to grade and to troubleshoot.

As we shall see below, the method of velocity factors shares this
pedagogical virtue, while nonetheless bringing us most of the theoretical
and practical virtues of rapidity.

\section{Justification of the method of velocity factors}

Define the velocity factor $f$ corresponding to $V$ by\begin{equation}
f=g(V)=\frac{1+V}{1-V}.\label{eq:velocity-factor-definition}\end{equation}

We note in passing that $g$ is a M\"obius function that rotates
the Riemann sphere by a quarter turn, with fixed points $\pm i;$
since all the coefficients are real, the real axis maps to itself.
In particular, if one stereographically projects the the real axis
onto a unit circle centred at 0, then $g$ corresponds to a quarter
turn of this circle, taking -1 to 0, 0 to 1, 1 to $\pm\infty$, and
$\pm\infty$ to -1. Composing $g$ twice yields the negative reciprocal
function, composing it three times yields its inverse, and composing
it four times yields the identity.

The connection between M\"obius functions and relativity proves remarkably
deep \cite{Net97}; this particular transformation has other computational
uses \cite{Dor93}. We shall not require these properties, though,
to prove what we need.

Solving for $V,$ we get

\begin{equation}
V=\frac{f-1}{f+1}.\label{eq:velocity-from-velocity-factor}\end{equation}

The correspondence between $V$ and $f$ is monotonic increasing,
with the velocity range $[-1,\;1]$ corresponding to the velocity
factor range $[0,\;\infty]$. $V=0$ corresponds to $f=1.$

Clearly, if $\bar{V}=-V$ then $\bar{f}=f^{-1};$ negation of velocities
corresponds to reciprocation of velocity factors.

Now

\begin{eqnarray}
f_{\textrm{ab}}\times f_{\textrm{bc}} & = & \frac{1+V_{\textrm{ab}}}{1-V_{\textrm{ab}}}\times\frac{1+V_{\textrm{bc}}}{1-V_{\textrm{bc}}}\nonumber \\
 & = & \frac{1+V_{\textrm{ab}}\cdot V_{\textrm{bc}}+V_{\textrm{ab}}+V_{\textrm{bc}}}{1+V_{\textrm{ab}}\cdot V_{\textrm{bc}}-V_{\textrm{ab}}-V_{\textrm{bc}}}\nonumber \\
 & = & \left(\frac{1+V_{\textrm{ab}}\cdot V_{\textrm{bc}}+(V_{\textrm{ab}}+V_{\textrm{bc}})}{1+V_{\textrm{ab}}\cdot V_{\textrm{bc}}}\right)\nonumber \\
 &  & \div\left(\frac{1+V_{\textrm{ab}}\cdot V_{\textrm{bc}}-(V_{\textrm{ab}}+V_{\textrm{bc}})}{1+V_{\textrm{ab}}\cdot V_{\textrm{bc}}}\right)\nonumber \\
 & = & \frac{1+(V_{\textrm{ab}}\boxplus V_{\textrm{bc}})}{1-(V_{\textrm{ab}}\boxplus V_{\textrm{bc}})}\nonumber \\
 & = & \frac{1+V_{\textrm{ac}}}{1-V_{\textrm{ac}}}\nonumber \\
 & = & f_{\textrm{ac}}.\label{eq:multiplicativity-derivation}\end{eqnarray}

So relativistic addition of velocities corresponds to ordinary multiplication
of velocity factors.

This result might have been had more quickly from the connection with
rapidities,

\begin{eqnarray}
\alpha & = & \arctanh V\nonumber \\
 & = & \ln\sqrt{\frac{1+V}{1-V}}\nonumber \\
 & = & \frac{1}{2}\cdot\ln f\nonumber \\
 & = & \log_{(\textrm{e}^{2})}f.\label{eq:rapidity-is-log-of-velocity-factor}\end{eqnarray}

but the derivation in Eq.~(\ref{eq:multiplicativity-derivation})
does not require any acquaintance with either rapidities or hyperbolic
functions, or indeed logarithms, exponentials or calculus.

In our examples, we have also made use of the observation that if,
for any $N$ and $D,$ \begin{equation}
V=\frac{N}{D}\label{eq:velocity-quotient}\end{equation}

then \begin{equation}
f=\frac{D+N}{D-N},\label{eq:velocity-factor-as-sumdif}\end{equation}

and that conversely if \begin{equation}
f=\frac{\nu}{\delta}\label{eq:velocity-factor-as-ratio}\end{equation}

then

\begin{equation}
V=\frac{\nu-\delta}{\nu+\delta}.\label{eq:velocity-as-sumdif}\end{equation}

In particular, if either of $V$ or the corresponding velocity factor
$f$ is rational, or more generally algebraic, then both are. Converting
either way requires taking a sum and a difference and forming a ratio;
this determines the target value up to a possible sign change and
a possible reciprocation, both of which can easily be put in by hand
if montonicity and the following correspondences are remembered :

\begin{tabular}{c|c}
$V$&
$f$\tabularnewline
\hline
\hline
-1&
0\tabularnewline
\hline
0&
1\tabularnewline
\hline
1&
$\infty$\tabularnewline
\end{tabular}

An alternative mnemonic can be derived from the trigonometric subtraction
formula\[
\tan(\psi-\phi)=\frac{\tan(\psi)-\tan(\phi)}{1+\tan(\psi).\tan(\phi)}.\]
Let $\psi-\phi=\pi/4,$ so that the left hand side is unity. Then
we can solve for $\tan\psi$ to get\begin{equation}
\tan(\psi)=\frac{1+\tan(\phi)}{1-\tan(\phi)}\label{eq:tangent-mnemonic}\end{equation}
Comparing this with Eq.~(\ref{eq:velocity-factor-definition}) we
see that if a velocity $V$ and its corresponding velocity factor
$f$ are regarded as the slopes of two lines, than the line whose
slope is $f$ is rotated by $+45^{\circ}$ relative to the line whose
slope is $V.$

\section{Using velocity factors}

We now turn to the use of this multiplicative representation, in which
velocities between $-1$ and 1 become velocity factors between 0 and
$\infty.$ We shall see that this representation fits somewhere between
the velocity representation and the rapidity representation. This
correspondence too is a monotonic increasing, invertible function
of velocity, but here zero velocity corresponds to a velocity factor
of 1, and negation of a velocity corresponds to reciprocation of its
velocity factor.

\subsection*{A first example: adding three given velocities relativistically. }

Suppose, e.g., that we wish to find\[
\frac{1}{3}\boxplus\frac{2}{5}\boxplus\left(-\frac{1}{4}\right),\]

where we are taking $c=1.$

We make a table containing the velocities we wish to sum :

\begin{tabular}{c|c}
$V$&
$f$\tabularnewline
\hline
\hline
$\frac{1}{3}$&
$\dots$\tabularnewline
\hline
$\frac{2}{5}$&
$\dots$\tabularnewline
\hline
$\frac{-1}{4}$&
$\dots$\tabularnewline
\hline
$\dots$&
$\dots$\tabularnewline
\end{tabular}

Then we compute the corresponding velocity factors :

\begin{tabular}{c|c}
$V$&
$f$\tabularnewline
\hline
\hline
$\frac{1}{3}$&
$\frac{3+1}{3-1}=2$\tabularnewline
\hline
$\frac{2}{5}$&
$\frac{5+2}{5-2}=\frac{7}{3}$\tabularnewline
\hline
$\frac{-1}{4}$&
$\frac{4-1}{4+1}=\frac{3}{5}$\tabularnewline
\hline
$\dots$&
$\dots$\tabularnewline
\end{tabular}

The values of the velocity factor $f$ are computed by forming ratios
of the sum and the difference of the numerator and denominator of
the values of $V.$

Whether the sum or the difference should be in the numerator and what
sign the difference should carry are easily figured out by remembering
that the velocity factor cannot be negative, and that positive velocities
$V$ correspond to velocity factors greater than one. Another simple
mnemonic is derived below.

Next, we multiply the velocity factors we have found to get the overall
velocity factor :

\begin{tabular}{c|c}
$V$&
$f$\tabularnewline
\hline
\hline
$\frac{1}{3}$&
$\frac{3+1}{3-1}=2$\tabularnewline
\hline
$\frac{2}{5}$&
$\frac{5+2}{5-2}=\frac{7}{3}$\tabularnewline
\hline
$\frac{-1}{4}$&
$\frac{4-1}{4+1}=\frac{3}{5}$\tabularnewline
\hline
$\dots$&
$2\times\frac{7}{3}\times\frac{3}{5}=\frac{14}{5}$\tabularnewline
\end{tabular}

Finally, we form a ratio of the sum and difference of the denominator
and numerator of the overall velocity factor on the right to get the
velocity sum :

\begin{tabular}{c|c}
$V$&
$f$\tabularnewline
\hline
\hline
$\frac{1}{3}$&
$\frac{3+1}{3-1}=2$\tabularnewline
\hline
$\frac{2}{5}$&
$\frac{5+2}{5-2}=\frac{7}{3}$\tabularnewline
\hline
$\frac{-1}{4}$&
$\frac{4-1}{4+1}=\frac{3}{5}$\tabularnewline
\hline
$\frac{14-5}{14+5}=\frac{9}{19}$&
$2\times\frac{7}{3}\times\frac{3}{5}=\frac{14}{5}$\tabularnewline
\end{tabular}

Again, we need not memorize which way around to write the difference,
or whether to put it in the numerator or denominator. We need only
remember that velocity factors larger than one correspond to positive
velocities, and that the magnitude of a velocity can be no greater
than one. So, finally,\[
\frac{1}{3}\boxplus\frac{2}{5}\boxplus\left(-\frac{1}{4}\right)=\frac{9}{19}.\]

\subsection*{A second example: relativistic fractions}

What, relativistically, is 3/7 of $(5/8)c?$

Regarding $(5/8)c$ as the overall result of a large number of locally
equivalent small boosts, this question asks what the velocity is when
3/7 of these small boosts have been executed.

From another point of view, the question asks for the velocity of
a boost, which when repeated 7 times (in successive comoving frames),
gives the same result as boosting 3 times (again, in successive comoving
frames) by $(5/8)c.$ In other words, we want to find $U$ such that

\begin{equation}
U\boxplus U\boxplus U\boxplus U\boxplus U\boxplus U\boxplus U=\frac{5}{8}\boxplus\frac{5}{8}\boxplus\frac{5}{8}.\label{eq:7-reltimes-u-is3-reltimes-5over8}\end{equation}

Before resorting to velocity factors, let us try to solve this by
repeated use of the usual velocity addition formula on each side.
Then Eq.~(\ref{eq:7-reltimes-u-is3-reltimes-5over8}) can be written\begin{equation}
\frac{U^{7}+21\cdot U^{5}+35\cdot U^{3}+7\cdot U}{7\cdot U^{6}+35\cdot U^{4}+21\cdot U^{2}+1}=\frac{1085}{1112},\label{eq:Urat-is-1085-over-1112}\end{equation}
 or \begin{align}
1112\cdot U^{7}-7595\ \cdot U^{6}+23352\cdot U^{5}-37975\cdot U^{4}\nonumber \\
+38920\cdot U^{3}-22785\cdot U^{2}+7784\cdot U-1085 & =0.\label{eq:Upoly}\end{align}
Even exploiting the palindromic symmetry between the coefficients
in numerator and denominator of the left hand side of Eq.~(\ref{eq:Urat-is-1085-over-1112}),
finding a solution in closed form is non-trivial.

Using rapidities and a scientific calculator, a numerical answer is
easily obtained by evaluating\begin{equation}
\tanh\left(\frac{3}{7}\times\arctanh\left(\frac{5}{8}\right)\right)\approx0.3043.\label{eq:tanh-arctanh}\end{equation}
 Using velocity factors, we can produce a closed-form answer practically
by inspection.

We first take 5/8 and find its velocity factor, which is (5+8)/(8-5)
= 13/3.

We raise this to the 3/7 power to get $13^{3/7}/3^{3/7},$ the velocity
factor of the desired answer, and convert back to a velocity in closed
form,\begin{equation}
U=\frac{13^{3/7}-3^{3/7}}{13^{3/7}+3^{3/7}}.\label{eq:sumdiff-powers}\end{equation}

This result can be confirmed by evaluating the left hand side of Eq.~(\ref{eq:tanh-arctanh})
algebraically, as shown below in section \ref{sec:Appendix-1}.

It can also be confirmed from Eq.~(\ref{eq:7-reltimes-u-is3-reltimes-5over8})
as shown below in section \ref{sec:Appendix-2}. Using the velocity
addition formula to confirm a correct value already supplied takes
considerably more effort than finding that value using velocity factors.

\section{The velocity factor is a two-way doppler factor}

The velocity factor also has a simple physical interpretation as the
two-way doppler factor corresponding to a given separation speed.
Thus if $\mathcal{O}$ at rest at $x=0$ in a vacuum sends a light
pulse of duration $T$ to an mirror $\mathcal{M}$ travelling with
velocity $V$ along the positive $x$ axis, then then $\mathcal{O}$
will observe a reflected pulse to have a duration $f\cdot T.$ Moreover,
this suggests an obvious physical explanation of why velocity factors
compose by multiplication. Indeed, Bondi used multiplicative composition
as one of the postulates of his $k$-calculus\cite{Boh64}, an elegant
and accessible formulation of special relativity. His $k,$ a one-way
doppler factor, is our $f^{1/2}.$ Of course, any fixed nonzero power
of either of these would also be a faithful multiplicative representation.

A one-way doppler factor is simpler than a two-way doppler factor,
so Bondi's $k$ is simpler physically. Nonetheless, like rapidity,
Bondi's beautiful $k-$calculus has not become a part of standard
part of the pedagogy of introductory physics.

\section{Summary}

Velocity addition using the usual formula is unwieldy and of limited
usefulness. Rapidities are a more powerful, and easily applied to
a broader range of questions. The internal workings of that tool are,
like those of the calculators required to use them, usually left inaccessible.
This leads to difficulties in error checking and interpretation.

Velocity factors \emph{are} more or less the internal workings of
rapidity. The correspondence between velocity factors and velocities
is simple. The use of velocity factors places interesting questions
in easy reach, and so encourages tinkering; relativistic velocity
addition can now become a more rewarding part of the standard curriculum
than at present. Velocity factors make physical sense, provide closed
form answers, are at least as memorable as the usual velocity addition
formula and doppler formulae to which they are equivalent---and one
can use them without having to reach for a calculator.

\subsection*{Acknowledgments}

It is a pleasure to thank William Evenson, Ann Cox and Kent Harrison
for their comments on various drafts.

\bibliography{../../bibtex-bibliography}

\begin{thebibliography}{7}
\expandafter\ifx\csname natexlab\endcsname\relax\def\natexlab#1{#1}\fi
\expandafter\ifx\csname bibnamefont\endcsname\relax
  \def\bibnamefont#1{#1}\fi
\expandafter\ifx\csname bibfnamefont\endcsname\relax
  \def\bibfnamefont#1{#1}\fi
\expandafter\ifx\csname citenamefont\endcsname\relax
  \def\citenamefont#1{#1}\fi
\expandafter\ifx\csname url\endcsname\relax
  \def\url#1{\texttt{#1}}\fi
\expandafter\ifx\csname urlprefix\endcsname\relax\def\urlprefix{URL }\fi
\providecommand{\bibinfo}[2]{#2}
\providecommand{\eprint}[2][]{\url{#2}}

\bibitem[{\citenamefont{Bondi}(1964)}]{Boh64}
\bibinfo{author}{\bibfnamefont{H.}~\bibnamefont{Bondi}},
  \emph{\bibinfo{title}{Relativity and Common Sense: a New Approach to
  Einstein}} (\bibinfo{publisher}{Doubleday and Co.}, \bibinfo{address}{Garden
  City, NY}, \bibinfo{year}{1964}).

\bibitem[{\citenamefont{Bondi}(1965)}]{Boh64a}
\bibinfo{author}{\bibfnamefont{H.}~\bibnamefont{Bondi}}, in
  \emph{\bibinfo{booktitle}{Lectures on General Relativity and Gravitation:
  {B}randeis 1964 {S}ummer {I}nstitute in {T}heoretical {P}hysics}}, edited by
  \bibinfo{editor}{\bibfnamefont{A.}~\bibnamefont{Trautman}},
  \bibinfo{editor}{\bibfnamefont{F.~A.~E.} \bibnamefont{Pirani}},
  \bibnamefont{and} \bibinfo{editor}{\bibfnamefont{H.}~\bibnamefont{Bondi}}
  (\bibinfo{publisher}{Prentice Hall}, \bibinfo{address}{NJ},
  \bibinfo{year}{1965}), vol.~\bibinfo{volume}{1}, pp.
  \bibinfo{pages}{375--459}.

\bibitem[{\citenamefont{{Misner} et~al.}(1973)\citenamefont{{Misner}, {Thorne},
  and {Wheeler}}}]{MiThWh73}
\bibinfo{author}{\bibfnamefont{C.~W.} \bibnamefont{{Misner}}},
  \bibinfo{author}{\bibfnamefont{K.~S.} \bibnamefont{{Thorne}}},
  \bibnamefont{and} \bibinfo{author}{\bibfnamefont{J.~A.}
  \bibnamefont{{Wheeler}}}, \emph{\bibinfo{title}{{Gravitation}}}
  (\bibinfo{publisher}{San Francisco: W.H.~Freeman and Co.},
  \bibinfo{year}{1973}).

\bibitem[{\citenamefont{{M}arc L{\'e}vy-{L}eblond}(1980)}]{Lej80}
\bibinfo{author}{\bibfnamefont{J.}~\bibnamefont{{M}arc L{\'e}vy-{L}eblond}},
  \bibinfo{journal}{American Journal of Physics} \textbf{\bibinfo{volume}{48}},
  \bibinfo{pages}{345} (\bibinfo{year}{1980}).

\bibitem[{\citenamefont{Wilson}(in preparation)}]{Wia07musrel}
\bibinfo{author}{\bibfnamefont{A.~T.} \bibnamefont{Wilson}},
  \bibinfo{journal}{American Journal of Physics}  (\bibinfo{year}{in
  preparation}).

\bibitem[{\citenamefont{Needham}(1997)}]{Net97}
\bibinfo{author}{\bibfnamefont{T.}~\bibnamefont{Needham}},
  \emph{\bibinfo{title}{Visual Complex Analysis}} (\bibinfo{publisher}{Oxford
  University Press}, \bibinfo{address}{Oxford}, \bibinfo{year}{1997}),
  \bibinfo{edition}{2nd} ed.

\bibitem[{\citenamefont{Doerfler}(1993)}]{Dor93}
\bibinfo{author}{\bibfnamefont{R.~W.} \bibnamefont{Doerfler}},
  \emph{\bibinfo{title}{Dead Reckoning: Calculating Without Instruments}}
  (\bibinfo{publisher}{Gulf Publishing Company}, \bibinfo{address}{Houston,
  TX}, \bibinfo{year}{1993}).

\end{thebibliography}

\section{Solution of relativistic fraction of velocity using rapidities \label{sec:Appendix-1}}

Evaluating expression (\ref{eq:tanh-arctanh}) we find

\begin{eqnarray*}
\tanh(\frac{3}{7}\times\arctanh(\frac{5}{8}))\quad\quad\end{eqnarray*}

\begin{eqnarray*}
 & = & \tanh(\frac{3}{7}\times\ln\sqrt{\frac{1+(5/8)}{1-(5/8)}})\\
 & = & \tanh(\frac{3}{14}\times\ln\frac{8+5}{8-5})\\
 & = & \tanh(\ln((\frac{13}{3})^{3/14})\\
 & = & \frac{\exp(\ln((13/3)^{3/14}))-\exp(-\ln((13/3)^{3/14}))}{\exp(\ln((13/3)^{3/14}))+\exp(-\ln((13/3)^{3/14}))}\\
 & = & \frac{(13/3)^{3/14}-(13/3)^{-3/14}}{(13/3)^{3/14}+(13/3)^{-3/14}}\\
 & = & \frac{(13/3)^{3/7}-1}{(13/3)^{3/7}+1}\\
 & = & \frac{13^{3/7}-3^{3/7}}{13^{3/7}+3^{3/7}},\end{eqnarray*}

as claimed in Eq.~(\ref{eq:sumdiff-powers}).

\section{solution of relativistic fraction of velocity using the usual velocity
addition formula \label{sec:Appendix-2}}

Evaluating the first relativistic addition on the right hand side
of Eq.~(\ref{eq:7-reltimes-u-is3-reltimes-5over8}) we find

\[
\frac{5}{8}\boxplus\frac{5}{8}=\frac{2(5/8)}{1+(5/8)^{2}}=\frac{80}{89},\]
so the whole right hand side of Eq.~(\ref{eq:7-reltimes-u-is3-reltimes-5over8})
becomes\begin{eqnarray}
\frac{5}{8}\boxplus\frac{5}{8}\boxplus\frac{5}{8} & = & \frac{80}{89}\boxplus\frac{5}{8}\nonumber \\
 & = & \frac{(80/89)+(5/8)}{1+(80/89)(5/8)}\nonumber \\
 & = & \frac{1085}{1112}.\label{eq:rhs-is-1085-over-1112}\end{eqnarray}

It is in principle straightforward to evaluate the left hand side
of Eq.~(\ref{eq:7-reltimes-u-is3-reltimes-5over8}),\begin{equation}
U\boxplus U\boxplus U\boxplus U\boxplus U\boxplus U\boxplus U\label{eq:7relU}\end{equation}
with\begin{equation}
U=\frac{13^{3/7}-3^{3/7}}{13^{3/7}+3^{3/7}}.\label{eq:Usoln}\end{equation}
 Evaluating expression (\ref{eq:7relU}) gives\begin{equation}
\frac{U^{7}+21\cdot U^{5}+35\cdot U^{3}+7\cdot U}{7\cdot U^{6}+35\cdot U^{4}+21\cdot U^{2}+1}.\label{eq:Urational-fun}\end{equation}
Substituting the right hand side of Eq.~(\ref{eq:Usoln}) into this
is tedious, even when one exploits the symmetries in both expressions.

We can instead reduce the labor by considering the expression

\begin{equation}
\frac{p^{m}-q^{m}}{p^{m}+q^{m}}\boxplus\frac{p^{n}-q^{n}}{p^{n}+q^{n}}\label{eq:pqmn-boxsum}\end{equation}

with arbitrary positive $p$ and $q$ and arbitrary real $m$ and
$n.$

Expanding the relativistic sum, we get\begin{equation}
\frac{\frac{p^{m}-q^{m}}{p^{m}+q^{m}}+\frac{p^{n}-q^{n}}{p^{n}+q^{n}}}{1+(\frac{p^{m}-q^{m}}{p^{m}+q^{m}})(\frac{p^{n}-q^{n}}{p^{n}+q^{n}})}.\label{eq:pqmn-boxsum-expanded}\end{equation}
Multiplying numerator and denominator by $(p^{m}+q^{m})(p^{n}+q^{n}),$
this becomes\begin{equation}
\frac{(p^{m}-q^{m})(p^{n}+q^{n})+(p^{n}-q^{n})(p^{m}+q^{m})}{(p^{m}+q^{m})(p^{n}+q^{n})+(p^{m}-q^{m})(p^{n}-q^{n})},\label{eq:pqmn-sumdif-prods}\end{equation}
 which expands to

\begin{widetext}\begin{equation}
\frac{p^{m+n}+p^{m}q^{n}-p^{n}q^{m}-q^{m+n}+p^{m+n}+p^{n}q^{m}-p^{m}q^{n}-q^{m+n}}{p^{m+n}+p^{m}q^{n}+p^{n}q^{m}+q^{m+n}+p^{m+n}-p^{m}q^{n}-p^{n}q^{m}+q^{m+n}}\label{eq:pqmn-sumdif-prods-expanded}\end{equation}

\end{widetext}

or\begin{equation}
\frac{2p^{m+n}-2q^{m+n}}{2p^{m+n}+2q^{m+n}}.\label{eq:pqmn-sumdif-reduced}\end{equation}

Thus we have \begin{equation}
\frac{p^{m}-q^{m}}{p^{m}+q^{m}}\boxplus\frac{p^{n}-q^{n}}{p^{n}+q^{n}}=\frac{p^{m+n}-q^{m+n}}{p^{m+n}+q^{m+n}}.\label{eq:p-q-power-addition}\end{equation}
 (This result could have been had at once from\[
(\frac{p}{q})^{m}\times(\frac{p}{q})^{n}=(\frac{p}{q})^{m+n},\]
 in which the two factors on the left and the product on the right
are each taken to be velocity factors.)

We now apply this general result (\ref{eq:p-q-power-addition}) to
our problem. Taking \begin{equation}
U=\frac{p-q}{p+q}=\frac{p^{1}-q^{1}}{p^{1}+q^{1}}\label{eq:Upq}\end{equation}
 and applying Eq.~(\ref{eq:p-q-power-addition}) repeatedly, it should
be clear that\begin{equation}
U\boxplus U\boxplus U\boxplus U\boxplus U\boxplus U\boxplus U=\frac{p^{7}-q^{7}}{p^{7}+q^{7}}.\label{eq:Upq-iter}\end{equation}

Now setting $p=13^{3/7}$ and $q=3^{3/7},$ the left hand side of
Eq.~(\ref{eq:7-reltimes-u-is3-reltimes-5over8}) becomes\begin{eqnarray}
U\boxplus U\boxplus U\boxplus U\boxplus U\boxplus U\boxplus U & = & \frac{(13^{3/7})^{7}-(3^{3/7})^{7}}{(13^{3/7})^{7}+(3^{3/7})^{7}}\nonumber \\
 & = & \frac{13^{3}-3^{3}}{13^{3}+3^{3}}\nonumber \\
 & = & \frac{2170}{2224}\nonumber \\
 & = & \frac{1085}{1112},\label{eq:U-relsum-result}\end{eqnarray}
 which is what Eq.~(\ref{eq:rhs-is-1085-over-1112}) gave us for
the right hand side of Eq.~(\ref{eq:7-reltimes-u-is3-reltimes-5over8}),
so that they are equal as claimed.
\end{document}